# Dimension Reduction of Health Data Clustering

Rahmat Widia Sembiring[1], Jasni Mohamad Zain[2], Abdullah Embong[3]
[1,2]Faculty of Computer System and Software Engineering,
Universiti Malaysia Pahang
Lebuhraya Tun Razak, 26300, Kuantan, Pahang Darul Makmur, Malaysia
[3]School of Computer Science, Universiti Sains Malaysia
11800 Minden, Pulau Pinang, Malaysia
[1]rahmatws@yahoo.com, [2]jasni@ump.edu.my, [3]ae@cs.usm.my

**ABSTRACT**

The current data tends to be more complex than conventional data and need dimension reduction. Dimension reduction is important in cluster analysis and creates a smaller data in volume and has the same analytical results as the original representation. A clustering process needs data reduction to obtain an efficient processing time while clustering and mitigate curse of dimensionality. This paper proposes a model for extracting multidimensional data clustering of health database. We implemented four dimension reduction techniques such as Singular Value Decomposition (SVD), Principal Component Analysis (PCA), Self Organizing Map (SOM) and FastICA. The results show that dimension reductions significantly reduce dimension and shorten processing time and also increased performance of cluster in several health datasets.

## 1 Introduction

The current data tends to be multidimensional and high dimension, and more complex than conventional data. Many clustering algorithms have been proposed and often produce clusters that are less meaningful. The use of multidimensional data will result in more noise, complex data, and the possibility of unconnected data entities. This problem can be solved by using clustering algorithm. Several clustering algorithms grouped into cell-based clustering, density based clustering, and clustering oriented. To obtain an efficient processing time to mitigate a curse of dimensionality while clustering, a clustering process needs data reduction. Dimension reduction is a technique that is widely used for various applications to solve curse of dimensionality.

Dimension reduction is important in cluster analysis, which not only makes the high dimensional data addressable and reduces the computational cost, but can also provide users with a clearer picture and visual examination of the data of interest [6]. Many emerging dimension reduction techniques proposed, such as Local Dimensionality Reduction (LDR) tries to find local correlations in the data, and performs dimensionality reduction on the locally correlated clusters of data individually [3], where dimension reduction as a dynamic process adaptively adjusted and integrated with the clustering process [4].

Sufficient Dimensionality Reduction (SDR) is an iterative algorithm [8],

which converges to a local minimum of $p^* = \arg \min_{\tilde{p} \in P\theta} D_{KL}[p|\tilde{p}]$ and hence solves the Max-Min problem as well. A number of optimizations can solve this minimization problem, and reduction algorithm based on Bayesian inductive cognitive model used to decide which dimensions are advantageous [11]. Developing an effective and efficient clustering method to process multidimensional and high dimensional dataset is a challenging problem.

This paper is organized into a few sections. Section 2 will present the related work. Section 3 explains the materials and method. Section 4 elucidates the results followed by discussion in Section 5. Section 6 deals with the concluding remarks.

## 2 Related Work

Functions of data mining are association, correlation, prediction, clustering, classification, analysis, trends, outliers and deviation analysis, and similarity and dissimilarity analysis. Clustering technique is applied when there is no class to predict but rather when the instances divide into natural groups [20]. Clustering for multidimensional data has many challenges. These are noise, complexity of data, and data redundancy. To mitigate these problems dimension reduction needed. In statistics, dimension reduction is the process of reducing the number of random variables. The process classified into feature selection and feature extraction [18], and the taxonomy of dimension reduction problems [16] shown in Figure.1. Dimension reduction is the ability to identify a small number of important inputs (for predicting the target) from a much larger number of available inputs, and is effective in cases when there are more inputs than cases or observations.

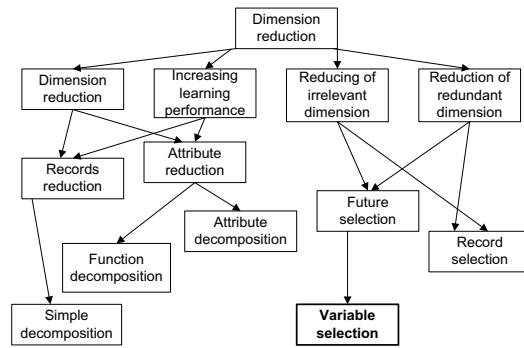

**Figure 1.** Taxonomy of dimension reduction problem

Dimension reduction methods associated with regression, additive models, neural network models, and methods of Hessian [6], one of which is the local dimension reduction (LDR), which is looking for relationships in the dataset and reduce the dimensions of each individual, then using a multidimensional index structure [3]. Nonlinear algorithm gives better performance than PCA for sound and image data [14], on the other studies mentioned Principal Component Analysis (PCA) is based on dimension reduction and texture classification scheme can be applied to manifold statistical framework [3].

In most applications, dimension reduction performed as pre-processing step [5], performed with traditional statistical methods that will parse an increasing number of observations [6]. Reduction of dimensions will create a more effective domain characterization [1]. Sufficient Dimension Reduction (SDR) is a generalization of nonlinear regression problems, where the extraction of features is as important as the matrix factorization [8], while SSDR (Semi-supervised Dimension Reduction) is used to maintain the original structure of high dimensional data [27].

The goals of dimension reduction methods are to reduce the number of predictor components and to help ensure that these components are independent. The method designed to provide a framework for interpretability of the results, and to find a mapping F that maps the input data from the space $\Re^d$ to lower dimension feature space $\Re^d$ denotes as $F(x): \Re^d \to \Re^{d'}$ [26, 15]. Dimension reduction techniques, such as principal component analysis (PCA) and partial least squares (PLS) can used to reduce the dimension of the microarray data before certain classifier is used [25].

We compared four dimension reduction techniques and embedded in DBSCAN, these dimension reduction are:

A. SVD

The Singular Value Decomposition (SVD) is a factorization of a real or complex matrix. The equation for SVD of *X* is $X=USV^T$ [24], where *U* is an *m* x *n* matrix, *S* is an *n* x *n* diagonal matrix, and $V^T$ is also an *n* x *n* matrix. The columns of *U* are called the *left singular vectors*, {$u_k$}, and form an orthonormal basis for the assay expression profiles, so that $\mathbf{u}_i \cdot \mathbf{u}_j = 1$ for *i* = *j*, and $u_i \cdot u_j = 0$ otherwise. The rows of $V^T$ contain the elements of the *right singular vectors*, {$\mathbf{v}_k$}, and form an orthonormal basis for the gene transcriptional responses. The elements of *S* are only nonzero on the diagonal, and are called the *singular values*. Thus, $S = \text{diag}(s_1,...,s_n)$. Furthermore, $s_k > 0$ for $1 \leq k \leq r$, and $s_i = 0$ for $(r+1) \leq k \leq n$. By convention, the ordering of the singular vectors is determined by high-to-low sorting of singular values, with the highest singular value in the upper left index of the *S* matrix.

B. PCA

PCA is a dimension reduction technique that uses variance as a measure of interestingness and finds orthogonal vectors (principal components) in the feature space that accounts for the most variance in the data [19]. Principal component analysis is probably the oldest and best known of the techniques of multivariate analysis, first introduced by Pearson, and developed independently by Hotelling [12].

The advantages of PCA are identifying patterns in data, and expressing the data in such a way as to highlight their similarities and differences. It is a powerful tool for analysing data by finding these patterns in the data. Then compress them by dimensions reduction without much loss of information [23]. Algorithm PCA [7] shown as follows:

a. Recover basis:
   Calculate $XX^T = \sum_{i=1}^{t} x_i x_i^T$ and let $U$ = eigenvectors of $XX^T$ corresponding to the top *d* eigenvalues.
b. Encode training data:
   $Y = U^T X$ where *Y* is a *d* x *t* matrix of encodings of the original data.
c. Reconstruct training data:
   $\hat{X} = UY = UU^T X$
d. Encode test example:
   $y = U^T > X$ where *y* is a *d*-dimensional encoding of *x*.
e. Reconstruct test example:
   $\hat{x} = U_y = UU^T x$

C. SOM

A self-organizing map (SOM) is a type of artificial neural network that is trained using unsupervised learning to produce a low-dimensional (typically two-dimensional), discretized representation of the input space of the training samples, and called a map [14]. Self-organizing maps are different from other artificial neural networks in the sense that they use a neighborhood function to preserve the topological properties of the input space. This makes SOMs useful for visualizing low-dimensional views of high-dimensional data, akin to multidimensional scaling. The model was first described as an artificial neural network by the Finnish professor Teuvo Kohonen, and is sometimes called a Kohonen map.

D. FastICA

Independent Component Analysis (ICA) introduced by Jeanny Hérault and Christian Jutten in 1986, later clarified by Pierre Comon in 1994 [22]. FastICA is one of the extensions of ICA, which is based on point iteration scheme to find the nongaussianity $w^T x$ [9], can also be derived as approximate Newton iteration, FastICA using the following formula:

$$W^+ = W + diag(\alpha_i)[diag(\beta_i) + E\{g(y)y^T\}]W$$

where $y = W_x, \beta_i = -E\{y_i g(y_i)\}$ and $\alpha_i = -1/(\beta_i - E\{g'(y_i)\})$, matrices $W$ need to *orthogonalized* after each phase have been processed.

## 3 Material and Method

This study is designed to find the most efficient dimension reduction technique. In order to achieve this objective, we implemented a model for efficiency of the cluster performed by first reducing the dimensions of datasets [21]. There are four dimension reduction techniques tested in the proposed model, namely SVD, PCA, SOM, FastICA.

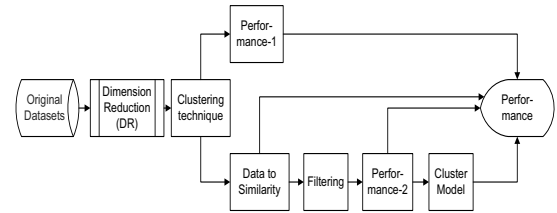

**Figure 2.** Proposed model compared based on dimension reduction and DBSCAN clustering

Dimensions reduction result is processed into DBSCAN cluster technique. DBSCAN needs *ε (eps)* and the minimum number of points required to form a cluster (*minPts*) including mixed euclidean distance as distance measure. For the result of DBSCAN clustering using functional data to similarity, it calculates a similarity measure from the given data (attribute based), and another output of DBSCAN that is measured is *performance-1*, this simply provides the number of clusters as a value.

Result of *data to similarity* takes an exampleSet as input for filter examples and returns a new exampleSet including only the examples that fulfill a condition. By specifying an implementation of a condition, and a parameter string, arbitrary filters can be applied and directly derive a *performance-2* as measure from a specific data or statistics value, then process expectation maximum cluster with parameter *k*=2, *max runs*=5, *max optimization step*=100, *quality*=1.0E-10 and *install distribution=k-means* run.

## 4 Result

Testing of model performance was conducted on four datasets model; e-coli, acute implant, blood transfusion and prostate cancer. Dimension reduction used SVD, PCA, SOM and FastICA. By using RapidMiner, we conducted the testing process without dimension reduction and clustering, and then compared with the results of clustering process using dimension reduction. Result of attribute dimension reduction shown in Table 1.

**Table 1.** Attribute Dimension Reduction

| Dimension reduction | Number of attribute for each datasets ||||
| --- | --- | --- | --- | --- |
| | E-coli | Acute implant | Blood transfusion | Prostate cancer |
| with SVD | 1 | 1 | 1 | 1 |
| with PCA | 5 | 4 | 1 | 3 |
| with SOM | 2 | 2 | 1 | 2 |
| with FastICA | 8 | 8 | 5 | 18 |
| without dimension reduction | 8 | 8 | 5 | 18 |

To find out efficiency we conducted the testing and record for processing time, as shown in Table 2.

**Table 2.** Processing time

| Dimension reduction | Processing time for each datasets ||||
| --- | --- | --- | --- | --- |
| | E-coli | Acute implant | Blood transfusion | Prostate cancer |
| with SVD | 19 | 9 | 61 | 39 |
| with PCA | 27 | 14 | 47 | 35 |
| with SOM | 34 | 22 | 51 | 41 |
| with FastICA | 67 | 12 | 58 | 148 |
| without dimension reduction | 22 | 11 | 188 | 90 |

Using SVD, PCA, SOM and FastICA we also conducted the testing process and found performance no of cluster, as shown in Table 3.

**Table 3.** Performance no of cluster

| Dimension reduction | Performance no of cluster for each datasets ||||
| --- | --- | --- | --- | --- |
| | E-coli | Acute implant | Blood transfusion | Prostate cancer |
| with SVD | 2 | 10 | 13 | 1 |
| with PCA | 2 | 2 | 2 | 2 |
| with SOM | 2 | 7 | 17 | 1 |
| with FastICA | 1 | 1 | 51 | 2 |
| without dimension reduction | 8 | 10 | 13 | 1 |

By implementing four different reduction techniques SVD, PCA, SOM, and FastICA, and continuously applying the cluster method based on cluster density. We obtained results, Figure. 3a present e-coli datasets based on DBSCAN without dimension reduction, Figure. 3b is the result of the cluster of E-coli datasets based on DBSCAN within SVD, Figure. 3c is a cluster of E-coli datasets based on DBSCAN and PCA, Figure. 3d is a cluster of E-coli datasets based on DBSCAN and SOM, while Figure. 3e is the result of the cluster by using DBSCAN within FastICA dimension reduction.

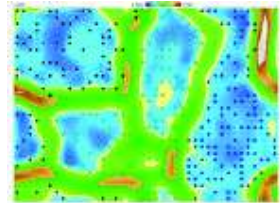
**Figure 3.1.** E-coli datasets based on DBSCAN without dimension reduction

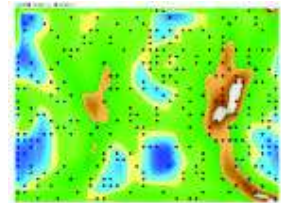
**Figure. 3b.** E-coli datasets based on DBSCAN and SVD

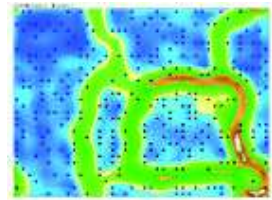
**Figure. 3c.** E-coli datasets based on DBSCAN and PCA

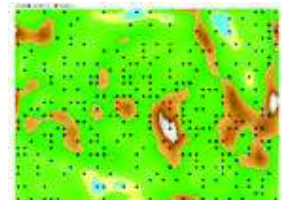
**Figure 3d.** E-coli datasets based on DBSCAN and SOM

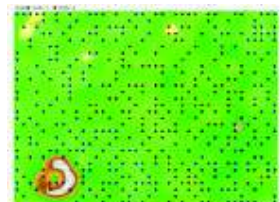
**Figure. 3e.** E-coli datasets based on DBSCAN and FastICA

Another result obtained for acute implant datasets, at Figure. 4a-e present acute implant datasets based on DBSCAN without dimension reduction and within a various dimension reduction.

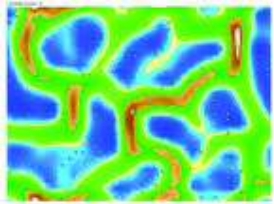

**Figure 4a.** Acute implant datasets based on DBSCAN without dimension reduction

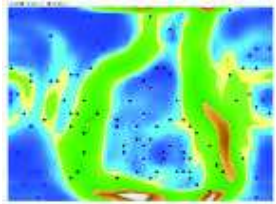

**Figure 4b.** Acute implant datasets based on DBSCAN and SVD

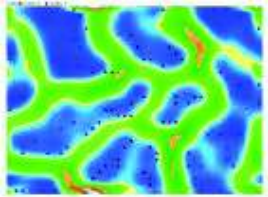

**Figure 4c.** Acute implant datasets based on DBSCAN and PCA

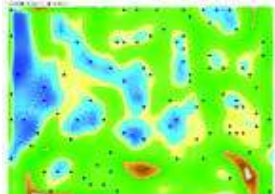

**Figure 4d.** Acute implant datasets based on DBSCAN and SOM

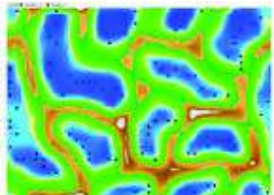

**Figure 4e.** Acute implant datasets based on DBSCAN and FastICA

The third was dataset tested is blood transfusion. Some of the result we present at Figure 5a-e, result obtained for blood transfusion datasets, based on DBSCAN without dimension reduction and within a various dimension reduction.

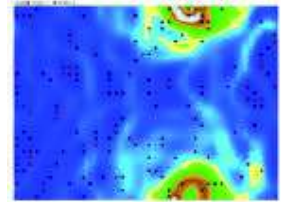

**Figure 5a.** Blood transfusion datasets based on DBSCAN without dimension reduction

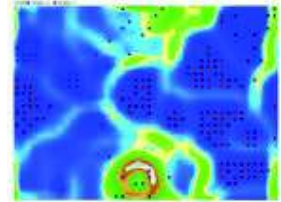

**Figure 5b.** Blood transfusion datasets based on DBSCAN and SVD

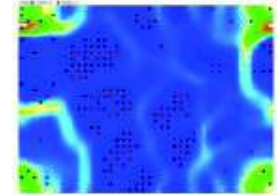

**Figure 5c.** Blood transfusion datasets based on DBSCAN and PCA

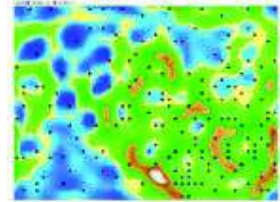

**Figure 5d.** Blood transfusion datasets based on DBSCAN and and SOM

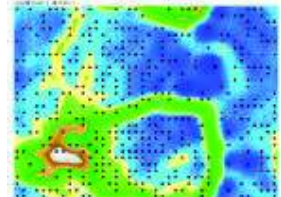

**Figure 5e.** Blood transfusion datasets based on DBSCAN and FastICA

Using same dimension reduction techniques, we clustered prostate cancer, result we present at Figure 6a-e, based on DBSCAN without dimension reduction and within a various dimension reduction.

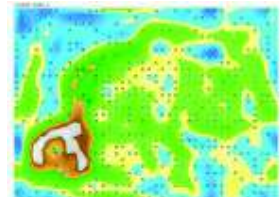

**Figure 6a.** Prostate cancer datasets based on DBSCAN without dimension reduction

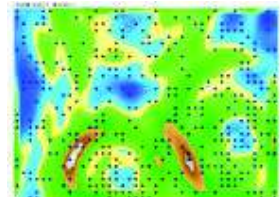

**Figure 6b.** Prostate cancer datasets based on DBSCAN and SVD

1046

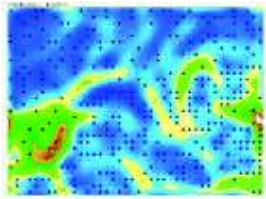

**Figure 6c.** Prostate cancer datasets based on DBSCAN and PCA

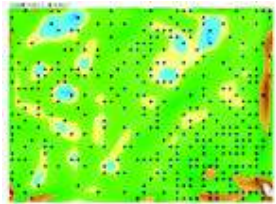

**Figure 6d.** Prostate cancer datasets based on DBSCAN and SOM

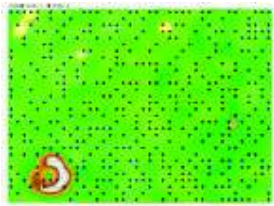

**Figure 6e.** Prostate cancer datasets based on DBSCAN and FastICA

Each cluster process, especially ahead of determined value of $\varepsilon=1$, and the value $MinPts=5$, while the number of clusters ($k=2$) that will be produced was also determined before.

## 5 Discussion

Dimension reduction before clustering process is to obtain efficient processing time and increase accuracy of cluster performance. Based on results in previous section, dimension reduction can shorten processing time and has lowest number of attribute. Figure. 7 shows DBSCAN with SVD has lowest number of reduced attribute.

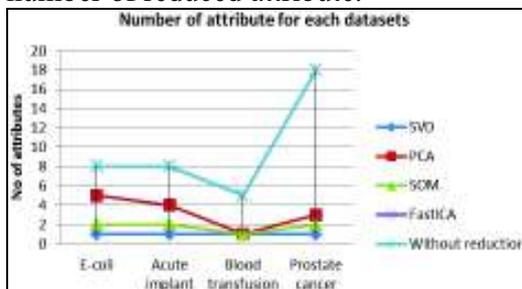

**Figure 7.** Reduction number of attributes

Another evaluation for model implementation is comparison of processing time. In general dimension reduction decreased time to process. For several datasets we found DBSCAN within SVD has lowest processing time.

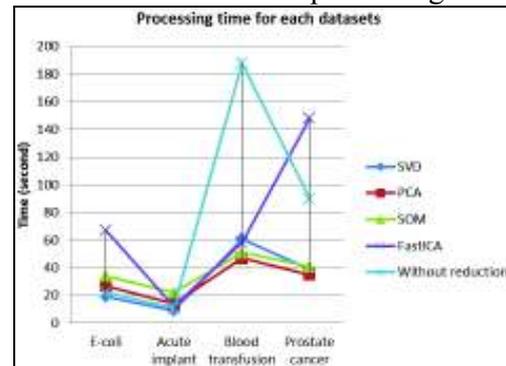

**Figure 8.** Processing time for each attribute

Cluster process with FastICA dimension reduction has highest cluster performance for blood datasets (Figure. 9), but lowest in other datasets, while PCA has lowest performance for overall datasets.

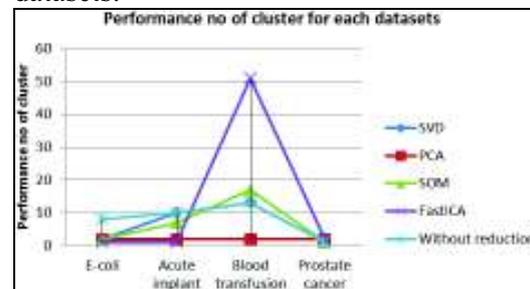

**Figure 9.** Performance no of cluster for each attribute

## 6 Conclusion

The discussion above has shown that applying a dimension reduction technique will shorten the processing time. Dimension reduction before clustering process is to obtain efficient processing time and increase cluster performance. DBSCAN with SVD has lowest processing time for several datasets. SVD also create lowest number of reduced attribute. In general,

1047

dimension reduction shows an increased cluster performance.